# Unsupervised Gene Expression Data using Enhanced Clustering Method

T.Chandrasekhar
Dept. of Computer Science
Periyar University
Salem, India
ch_ansekh80@rediffmail.com

K.Thangavel
Dept. of Computer Science
Periyar University
Salem, India
drktvel@gmail.com

E.Elayaraja
Dept. of Computer Science
Periyar University
Salem, India
elayarajaphd.e@gmail.com

E.N.Sathishkumar
Dept. of Computer Science
Periyar University
Salem, India
en.sathishkumar@yahoo.in

*Abstract*—Microarrays are made it possible to simultaneously monitor the expression profiles of thousands of genes under various experimental conditions. Identification of co-expressed genes and coherent patterns is the central goal in microarray or gene expression data analysis and is an important task in bioinformatics research. Feature selection is a process to select features which are more informative. It is one of the important steps in knowledge discovery. The problem is that not all features are important. Some of the features may be redundant, and others may be irrelevant and noisy. In this work the unsupervised Gene selection method and Enhanced Center Initialization Algorithm (ECIA) with K-Means algorithms have been applied for clustering of Gene Expression Data. This proposed clustering algorithm overcomes the drawbacks in terms of specifying the optimal number of clusters and initialization of good cluster centroids. Gene Expression Data show that could identify compact clusters with performs well in terms of the Silhouette Coefficients cluster measure.

## I. INTRODUCTION

Mining microarray gene expression data is an important research topic in bioinformatics with broad applications. Microarray technologies are powerful techniques for simultaneously monitoring the expression of thousands of genes under different sets of conditions. Gene expression data can be analyzed in two ways: unsupervised and supervised analysis. In supervised analysis, information about the structure/groupings of the object is assumed known or at least partially known. This prior knowledge is used in analysis process. In unsupervised analysis, prior knowledge is unknown.

Clustering of gene expression data can be divided into two main categories: Gene-based clustering and Sample-based clustering. In gene based clustering, genes are treated as objects and samples are features or attributes for clustering. The goal of gene-based clustering is to identify differentially expressed genes and sets of genes or conditions with similar expression pattern or profiles, and to generate a list of expression measurements. In Sample based clustering, samples are treated as objects and genes are features for clustering. Sample based clustering can be used to reveal the phenotype structure or substructure of samples. Applying the conventional clustering methods to cluster samples using all the genes as features may degrade the quality and reliability of clustering results.

Clustering has been used in a number of applications such as engineering, biology, medicine and data mining. Cluster analysis of gene expression data has proved to be a useful tool for identifying co-expressed genes. DNA microarrays are emerged as the leading technology to measure gene expression levels primarily, because of their high throughput. Results from these experiments are usually presented in the form of a data matrix in which rows represent genes and columns represent conditions or samples [13]. Each entry in the matrix is a measure of the expression level of a particular gene under a specific condition. Analysis of these data sets reveals genes of unknown functions and the discovery of functional relationships between genes [19]. Co-expressed genes can be grouped into clusters based on their expression patterns of gene based clustering and Sample based clustering. In gene based clustering, the genes are treated as the objects, while the samples are the features. In sample based clustering, the samples can be partitioned into homogeneous groups where the genes are regarded as features and the samples as objects [20].

Discriminant analysis is now widely used in bioinformatics, such as distinguishing cancer tissues from normal tissues or one cancer subtype vs. another [4]. A critical issue in discriminant analysis is feature selection: instead of using all available variables (features or attributes) in the data, one selectively chooses a subset of features to be used in the discriminant system. There are number of advantages of feature selection: (1) dimension reduction to reduce the computational cost; (2) reduction of noise to improve the classification accuracy; (3) more interpretable features or characteristics that can help identify and monitor the target diseases or function types. These advantages are typified in DNA microarray gene expression profiles. Of the tens of thousands of genes in experiments, only a smaller number of them show strong correlation with the targeted phenotypes [4, 17].

The most popular clustering algorithms in microarray gene expression analysis are Hierarchical clustering, K-Means clustering [3], and SOM [9]. Of these K-Means clustering is very simple and fast efficient. Numerous methods have been proposed to solve clustering problem. The most popular clustering methods are K-Means clustering algorithm which is developed by Mac Queen [7]. The K-Means algorithm is effective in producing clusters for many practical applications. But the computational complexity of the original K-Means algorithm is very high, especially for large data sets. The K-Means clustering algorithm is a partitioning clustering method that separates data into K groups. For the real life problems, the suitable number of clusters cannot be predicted. To overcome the above drawback the current research focused on developing the clustering algorithms without giving the initial number of clusters [2, 3, 5].

This paper is organized as follows. Section 2 presents and overview of Gene based clustering Techniques. Section 3 describes the Unsupervised Clustering algorithms. Section 4 describes performance of Experimental analysis and discussion. Section 5 conclusion and future work.

## II. GENE BASED CLUSTERING TECHNIQUES

The purpose of clustering gene expression data is to reveal the natural structure inherent data and extracting useful information from noisy data. The two class cancer subtype classification problem, 50 informative genes are usually sufficient. There are studies suggesting that only a few genes are sufficient [4]. Thus, computation is reduced while prediction accuracy is increased via effective feature selection. When a small number of genes are selected, their biological relationship with the target diseases is more easily identified. These "marker" genes provide additional scientific understanding of the problem [8]. Selecting an effective and more representative feature set is the subject of this paper.

### A. Analysis the Data

The gene expression data is min-max normalization by setting min 0 and max 1. Min-max normalization performs a linear transformation on the original data. Suppose that $min_A$ and $max_A$ are the minimum and maximum values of an attribute A. Min-max normalization maps a value of A to $v$ in the range [new_$min_A$, new_$max_A$] by computing

$$v' = \frac{v - min_A}{max_A - min_A}(new\_max_A - new\_min_A) + new\_min_A$$

Min-max normalization preserves the relationships among the original data values. It will encounter an "out-of-bounds" error if a future input case for normalization falls outside of the original data range for A. After the normalization of discretized value of gene $g_i$ at condition, $t_j$ is given the Gene Expression Data are normalizes then run on the discretized data. Discretization is then performed on this normalized expression data. The discretization is done as follows [4].

i. The discretized value of gene $g_i$ at condition, $t_1$ (i.e., the first condition)

$$\xi_{g_i,t_1} = \begin{cases} 1 & if\ \varepsilon_{g_i,t_j} > 0 \\ 0 & if\ \varepsilon_{g_i,t_1} = 0 \\ -1 & if\ \varepsilon_{g_i,t_1} < 0 \end{cases}$$

ii. The discretized values of gene $g_i$ at conditions $t_j$ ($j$ = 1, 2, ..., $T$ − 1) i.e., at the rest of the conditions ($T$ − $\{t_1\}$)

$$\xi_{g_i,t_{j+1}} = \begin{cases} 1 & if\ \varepsilon_{g_i,t_j} < \varepsilon_{g_i,t_{j+1}} \\ 0 & if\ \varepsilon_{g_i,t_j} = \varepsilon_{g_i,t_{j+1}} \\ -1 & if\ \varepsilon_{g_i,t_j} > \varepsilon_{g_i,t_{j+1}} \end{cases}$$

Where $\xi_{g_i,t_j}$ is the discretized value of gene $g_i$ at conditions $t_j$ ($j$ = 1, 2, ..., T − 1). The expression value of gene $g_i$ at condition $t_j$ is given by $g_i, t_j$. We see in the above computation that the first condition, $t_1$, is treated as a special case and its discretized value is directly based on $g_i, t_1$ i.e., the expression value at condition $t_1$. For the rest of the conditions the discretized value is calculated by comparing its expression value with that of the previous value. This helps in finding whether the gene is up 1 or down -1 regulated at that particular condition. Each gene will now have a regulation pattern of 0, 1, and -1 across the conditions or time points. This pattern is represented as a string.

### B. Unsupervised Gene Selection

Gene selection is an important problem in the research of Gene expression data analysis. In some cases, too many redundant or missing values are there in gene expression data. In this section, an existing works of Velayutham and Thangavel et al proposed an algorithm of unsupervised feature selection (Unsupervised Quick Reduct algorithms) [21] is applied. This method is based on dependency measure using rough set theory.

Rough set theory can be regarded as a new mathematical tool for imperfect data analysis. The theory has found applications in many domains. Objects characterized by the same information are indiscernible (similar) in view of the available information about them. The indiscernibility relation generated in this way is the mathematical basis of rough set theory. Any set of all indiscernible (similar) objects is called an elementary set, and forms a basic granule (atom) of knowledge about the universe [17, 24]. Any union of some elementary sets is referred to as a crisp (precise) set – otherwise the set is rough (imprecise, vague). Rough set is a pair of precise sets, called the lower and the upper approximation of the rough set, and is associated. The lower approximation consists of all objects which surely belong to the set and the upper approximation contains all objects which possibly belong to the set. The difference between the upper and the lower approximation constitutes the boundary region of the rough set. From a data table are called columns of which are labeled by attributes, rows – by objects of interest and entries of the table are attribute values.

In data mining applications, decision class labels are often unknown or incomplete. In this situation the unsupervised feature selection is play vital role to select features.

### C. Unsupervised Quick Reduct Algorithm[21]

The USQR algorithm attempts to calculate a reduct without exhaustively generating all possible subsets. It starts off with an empty set and adds in turn, one at a time, those attributes that result in the greatest increase in the rough set dependency metric, until this produces its maximum possible value for the dataset. According to the algorithm, the mean dependency of each attribute subset is calculated and the best candidate is chosen:

$$\gamma_R(\{a\}) = \frac{|POS_R(a)|}{|U|}, \forall a \in A$$

Where R is a reduct if and only if $K_R(\{a\})=K_c(\{a\})$ and $\forall x \subset R, \; Kx(\{a\}) \neq Kc(\{a\})$

Algorithm 1: The USQR algorithm

USQR (C)
C, the set of all conditional features;

(1) R ← { }
(2) do
(3) T ← R
(4)   ∀ x ∈ (C-R)
(5)   ∀ y ∈ C
(6)   $\gamma_{R \cup \{x\}}(y) = \frac{|POS_{R \cup \{x\}}(y)|}{|U|}$
(7)   if $\overline{\gamma_{R \cup \{x\}}(y)}, \forall y \in C > \overline{\gamma_T(y)}, \forall y \in C$
(8)     T ← R∪{x}
(9) R ← T
(10) until $\overline{\gamma_R(y)}, \forall y \in C = \overline{\gamma_C(y)}, \forall y \in C$
(11) return R

## III. UNSUPERVISED CLUSTERING ALGORITHMS

The reduced data selected by the gene selection algorithm USQR are clustered by using the K-Means algorithms.

### A. K-Means Clustering

The main objective in cluster analysis is to group objects that are similar in one cluster and separate objects that are dissimilar by assigning them to different clusters. One of the most popular clustering methods is K-Means clustering algorithm [3, 9, 12]. It is classifies objects to a pre-defined number of clusters, which is given by the user (assume *K* clusters). The idea is to choose random cluster centers, one for each cluster. These centers are preferred to be as far as possible from each other. In this algorithm mostly Euclidean distance is used to find distance between data points and centroids [6, 13, 22]. The Euclidean distance between two multi-dimensional data points $X = (x_1, x_2, x_3, ..., x_m)$ and $Y = (y_1, y_2, y_3, ..., y_m)$ is described as follows

$$D(X, Y) = \sqrt{(x_1 - y_1)^2 + (x_2 - y_2)^2 + \cdots + (x_m - y_m)^2}$$

The K-Means method aims to minimize the sum of squared distances between all points and the cluster centre. This procedure consists of the following steps, as described below.

Algorithm 2: K-Means clustering algorithm [18]

Require: $D = \{d_1, d_2, d_3, ..., d_n\}$ // Set of n data points.
    K - Number of desired clusters
Ensure: A set of K clusters.

Steps:
1. Arbitrarily choose *k* data points from *D* as initial centroids;
2. Repeat
    Assign each point $d_i$ to the cluster which has the closest centroid;
    Calculate the new mean for each cluster;
    Until convergence criteria is met.

Though the K-Means algorithm is simple, it has some drawbacks of quality of the final clustering, since it highly depends on the arbitrary selection of the initial centroids[1, 14].

### B. The Enhanced method

Performance of iterative clustering algorithms which converges to numerous local minima depends highly on initial cluster centers. Generally initial cluster centers are selected randomly. In this section, the cluster centre initialization algorithm is studied to improve the performance of the K-Means algorithm.

Algorithm 3: Enhanced Centre Initialization Algorithm (ECIA) [25]

Require: $D = \{d_1, d_2, d_3, ..., d_i, ..., d_n\}$ // Set of n data points.
    $d_i = \{x_1, x_2, x_3, ..., x_i, ..., x_m\}$ // Set of attributes of one data point.
    k // Number of desired clusters.
Ensure: A set of k clusters.

Steps:
1. In the given data set D, if the data points contains the both positive and negative attribute values then go to Step 2, otherwise go to step 4.
2. Find the minimum attribute value in the given data set D.
3. For each data point attribute, subtract with the minimum attribute value.
4. For each data point calculate the distance from origin.
5. Sort the distances obtained in step 4. Sort the data points accordance with the distances.
6. Partition the sorted data points into k equal sets.
7. In each set, take the middle point as the initial centroid.
8. Compute the distance between each data point $d_i$ ($1 \leq i \leq n$) to all the initial centroids $c_j$ ($1 \leq j \leq k$).
9. Repeat
10. For each data point $d_i$, find the closest centroid $c_j$ and assign $d_i$ to cluster j.
11. Set ClusterId[i]=j. // j:Id of the closest cluster.
12. Set NearestDist[i]= $d(d_i, c_j)$.
13. For each cluster j ($1 \leq j \leq k$), recalculate the centroids.
14. For each data point $d_i$,
14.1 Compute its distance from the centroid of the present nearest cluster.
14.2 If this distance is less than or equal to the present nearest distance, the data point stays in the same cluster.
    Else
14.2.1 For every centroid $c_j$ ($1 \leq j \leq k$) compute the distance $d(d_i, c_j)$.
    End for;
    Until convergence criteria is met.

## IV. EXPERIMENTAL ANALYSIS AND DISCUSSION

In this section, we describe the data sets used to analyze the methods studied in sections 2 and 3, which are arranged for the listed in Table 1, number of features/genes are in column wise, and number of items/samples are in row wise [23].

TABLE I. ARRANGE ALL THE GENE EXPRESSION DATA SETS

| Samples | Condition attributes(genes) | | | |
|---|---|---|---|---|
| | Gene 1 | Gene 2 | … | Gene n |
| 1 | g(1,1) | g(1,2) | … | g(1,q) |
| 2 | g(2,1) | g(2,2) | … | g(2,q) |
| … | … | … | … | … |
| p | g(p,1) | g(p,2) | … | g(p,q) |

*1) Serum data:* This data set is described and used in [10]. It can be downloaded from: http://www.sciencemag.org/ feature/ data/ 984559.shl and corresponds to the selection of 517 genes whose expression varies in response to serum concentration inhuman fibroblasts [11].

*2) Yeast data:* This data set is downloaded from Gene Expression Omnibus-databases. The Yeast cell cycle dataset contains 2884 genes and 17 conditions. To avoid distortion or biases arising from the presence of missing values in the data matrix we removal all the genes that had any missing value. This step results in a matrix of size 2882 * 17.

*3) Simulated data:* It is downloaded from http:// www.igbmc.ustrasbg.fr/ projects/fcm/y3c.txt. The set contains 300 Genes [3].

*4) Leukemia data:* It is downloaded from the website: http://datam.i2r.a-star.edu.sg/datasets/krbd/. The set contains 7129*34.

### A. Comparative Analysis

The ECIA with K-Means used to clustering in the all data sets after the section 2 as Table 2, the distance measure used here is the Euclidean distance. To access the quality of the clusters, we used the silhouette measure proposed by Rousseeuw [11, 15, 16]. In this method the Initial centroid values taken 7 then run as 10 times running to ECIA with K-means clusters data into 7 groups as Table 3.

TABLE II. COMPARATIVE ANALYSIS TO FILTER OF GENE EXPRESSION DATA

| Data sets | Before Unsupervised Gene Selection | After Unsupervised Gene selection |
|---|---|---|
| Serum | 517*17 | 49*17 |
| Yeast | 2884*17 | 118*17 |
| Simulated | 300*17 | 67*17 |
| Leukemia | 7129*34 | 142*34 |

TABLE III. COMPARATIVE ANALYSIS OF GENE EXPRESSION DATA

| Data sets Cluster | Serum (49*17) | | Yeast (118*17) | | Simulated (67*17) | | Leukemia (142*34) | |
|---|---|---|---|---|---|---|---|---|
| | No. of Gene | Measure value | No. of Gene | Measure value | No. of Gene | Measure value | No. of Gene | Measure value |
| C1 | 6 | 0.629 | 13 | 0.621 | 10 | 0.314 | 19 | 0.391 |
| C2 | 7 | 0.532 | 20 | 0.597 | 9 | 0.428 | 21 | 0.301 |
| C3 | 6 | 0.439 | 21 | 0.584 | 12 | 0.278 | 15 | 0.401 |
| C4 | 9 | 0.347 | 19 | 0.471 | 9 | 0.291 | 25 | 0.321 |
| C5 | 7 | 0.301 | 16 | 0.421 | 11 | 0.281 | 19 | 0.491 |
| C6 | 6 | 0.397 | 17 | 0.327 | 8 | 0.325 | 22 | 0.336 |
| C7 | 8 | 0.309 | 12 | 0.697 | 8 | 0.391 | 21 | 0.418 |
| Total Gene | 49 | | 118 | | 67 | | 142 | |

In table 3, the comparative analysis of gene expression data represent genes and measure value of four data sets. The Simulated data set produce high measure value comparing to other three data sets.

### B. Performance Comparison

In this various clusters performance is analysis to best cluster such as compact genes as Fig.1 and table 4. Among four data sets, each one produce compact cluster and provide best genes.

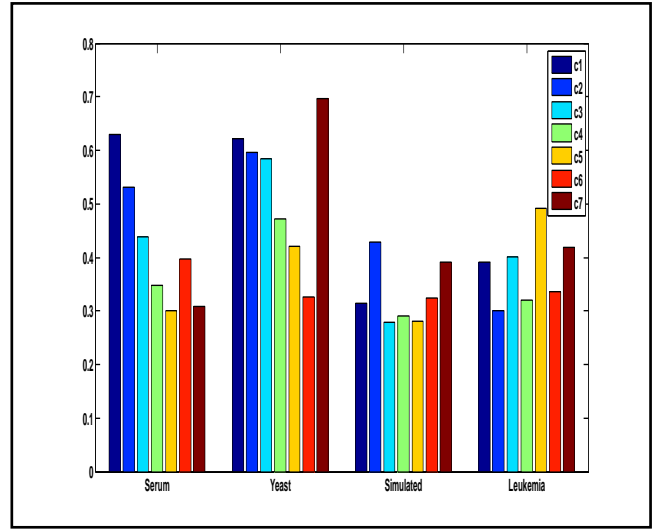

Fig. 1. Clusters Performance comparison chart for all data sets

TABLE IV. COMPARATIVE ANALYSIS OF GENE EXPRESSION DATA

| Data sets | Compact cluster | Number of Gene |
|---|---|---|
| Serum | C1 | 6 |
| Yeast | C7 | 12 |
| Simulated | C2 | 9 |
| Leukemia | C5 | 19 |

### V. CONCLUSION

In this work, unsupervised gene selections method USQR was studied and applies to avoid too many redundant or missing values in Microarray gene expression data. In this unsupervised gene selection method is based on unsupervised Feature selection using Rough set methods. These methods are used to get minimum number of random gene data sets and then we use K-Means clustering technique to improve the quality of clusters. One of the demerits of K-Means algorithm is random selection of initial seed point of desired clusters. This was overcome with ECIA for finding the initial centroids to avoid the random selection of initial values. Therefore, the ECIA algorithm is not depending upon any choice of the number of cluster and automatic evaluation of initial seed centroids and it produces different better results. Both the algorithms were tested with gene expression data and analysis the performance of cluster values using silhouette measurement. Therefore, finding solution to select the different centroids as clusters seed points and various measures are used to improve the cluster performance is our future endeavor.